\documentclass[aps,prd,onecolumn,amsmath,amsfonts,amssymb,groupedaddress,nofootinbib,nobalancelastpage,floatfix,secnumarabic,superscriptaddress,preprintnumbers]{revtex4-2}
\pdfoutput=1
\usepackage{epsfig, graphicx, hyperref, slashed,xspace}
\usepackage[dvipsnames]{xcolor}
\usepackage[utf8]{inputenc}
\usepackage[T1]{fontenc}
\usepackage{cleveref}
\usepackage{soul}
\usepackage{diagbox}
\usepackage{booktabs}
\usepackage{siunitx}
\usepackage{multirow}
\usepackage{setspace}
\usepackage{float}
\usepackage{lineno}

\usepackage[a4paper,left=1.1in,right=1.1in,top=1.2in,bottom=1.2in]{geometry}

\hypersetup{colorlinks=true,linkcolor=Maroon,citecolor=ForestGreen,filecolor=ForestGreen,urlcolor=ForestGreen}

\newcommand{\gev}{\ensuremath{\text{GeV}}\xspace}
\newcommand{\TeV}{\ensuremath{\text{TeV}}\xspace}

\newcommand{\MG}{\textsc{MadGraph5}\xspace}
\newcommand{\Pythia}{\textsc{Pythia8}\xspace}
\newcommand{\Powheg}{\textsc{Powheg-Box}\xspace}
\newcommand{\hvq}{\textsc{hvq}\xspace}
\newcommand{\bbfourl}{\textsc{bb4l}\xspace}
\newcommand{\MiNNLO}{\textsc{MiNNLO}$_{\text{PS}}$\xspace}
\newcommand{\Rivet}{\textsc{Rivet}\xspace}

\newcommand{\FastJet}{\textsc{FastJet}\xspace}

\newcommand{\MS}{\textsc{MadSpin}\xspace}
\newcommand{\LHAPDF}{\textsc{LhaPDF}\xspace}

\newcommand{\ATLASTwoFive}{\texttt{ATLAS2025}\xspace}
\newcommand{\CMSOneEight}{\texttt{CMS2018}\xspace}

\usepackage{xcolor}
\usepackage[normalem]{ulem}   

\makeatletter\def\l@subsubsection#1#2{}\makeatother
\usepackage{orcidlink}

\newcommand{\fittable}[1]{\resizebox{\ifdim\width>\linewidth \linewidth\else \width\fi}{!}{#1}}

\begin{document}

\title{Statistical analysis of the excess of events at the $t\bar{t}$ threshold at the LHC}

\author{Aman Desai \orcidlink{0000-0003-2631-9696}}\email{aman.desai@adelaide.edu.au}
\affiliation{Department of Physics, Adelaide University, Adelaide, SA 5005, Australia}

\author{Paul Jackson \orcidlink{0000-0002-0847-402X}}\email{p.jackson@adelaide.edu.au}
\affiliation{Department of Physics, Adelaide University, Adelaide, SA 5005, Australia}

\date{\today}
\begin{abstract}

\vspace{1mm}
\centerline{\bf \large Abstract}
\vspace{3mm}
The ATLAS and CMS experiments have reported an excess of events near the top-quark pair threshold relative to state-of-the-art perturbative QCD predictions, commonly attributed to the formation of a short-lived quasi-bound $t\bar{t}$ state (``toponium''). Several new Monte Carlo generator prescriptions, implemented in the \MG, \Pythia, and \Powheg frameworks, have since been developed to explain this excess. In this work, these new generator prescriptions are compared with published particle-level measurements available as \Rivet routines in the dilepton decay channel of top-quark pair production. The improvement offered by each model is quantified in terms of the $\chi^2$ agreement with the data for lepton-based angular and invariant dilepton mass observables sensitive to the threshold excess. It is found that including threshold contribution improves the $\chi^2/N_{\mathrm{dof}}$ relative to the corresponding baseline generator for most of the observables considered in this analysis.
\end{abstract}

\maketitle

\section{Introduction}\label{sec:intro}

Near the top-quark pair threshold ($m_{t\bar{t}} \sim 2 \cdot m_{t}$), the CMS and ATLAS experiments have recently reported an event yield in the dileptonic decay channel which is in excess of the perturbative QCD predictions~\cite{CMSThreshold2025,ATLASDiscovery2026}. In the CMS analysis, the excess is reported to have a cross section of $\sigma = 8.8^{+1.2}_{-1.4}$ pb~\cite{CMSThreshold2025}, while the ATLAS analysis reports a cross section of $\sigma = 9.3^{+1.4}_{-1.3}$ pb~\cite{ATLASDiscovery2026}. This observation has been interpreted as the formation of a short-lived quasi-bound state of a top quark and a top antiquark (``toponium'')~\cite{Fadin1987,Fadin1990,Strassler1991,Hagiwara2008,Kiyo2009,Sumino2010}. Owing to the large top-quark decay width, the resonance is not sharply peaked, in contrast to charmonium and bottomonium states. More recently, CMS has also reported an excess in the single-lepton decay channel of $\sigma = 5.1 \pm 0.9$ pb~\cite{CMSSingleLepton2026}, corroborating the effect independently of the dilepton channel.

At threshold, the top quarks move slowly relative to each other\footnote{The relative top velocity is given by $\beta_t=\sqrt{1-\frac{4m_t^{2}}{m_{t\bar{t}}^2}}$.}, and the dynamics are dominated by the exchange of multiple Coulomb-like gluons. This regime is described within non-relativistic QCD (NRQCD), in which the top-quark pair cross section is obtained from the corresponding non-relativistic Green's function~\cite{Strassler1991,Hagiwara2008}.  The observed excess is characterised as a pseudoscalar spin state. In addition to the spin properties, the $q\bar{q}\to t\bar{t}$ process proceeds via single-gluon exchange, so the resulting pair is in colour-octet configuration, whereas for the $gg\to t\bar{t}$, the process is a mixture of colour-singlet and colour-octet channels.

The CMS analysis~\cite{CMSThreshold2025} that reported the excess of events employed a simplified model~\cite{Fuks2021} in which the toponium resonance was modelled as a pseudoscalar state. Ref.~\cite{Fuks2025} subsequently introduced a new model, implemented in \MG~\cite{MG5AMC}, in which the top-quark pair matrix elements are reweighted by the non-relativistic Green's function. This model was used in the ATLAS study~\cite{ATLASDiscovery2026}.  This prescription is accurate to leading order (LO) in QCD. Several subsequent studies of the excess have used this model~\cite{Desai2026,Antozzi2026}.

In particular, Refs.~\cite{Fuks2026,ATLASThreshold2026} assess the significance of the excess by comparing Standard Model predictions without toponium formation with predictions in which toponium is included as an additional contribution to top-quark pair production. As their baseline predictions, Ref.~\cite{Fuks2026} has considered fixed-order NNLO calculations provided in Ref.~\cite{Czakon:2020qbd}. The study~\cite{ATLASThreshold2026} on the other hand considered the reweighted \Powheg models such as \hvq and \bbfourl samples along with \MiNNLO. In the above studies, the toponium modelling was carried out according to the \MG implementation of Ref.~\cite{Fuks2025}. In both studies it was found that data are explained better by including the bound state predictions in the analysis. 

More recently, additional prescriptions to simulate the top-quark pair threshold including the threshold enhancement effects have become available in \Pythia~\cite{PYTHIA83,Sjostrand2026} and in the \Powheg framework~\cite{Frixione2007_1,Oleari2010,Jezo2015,Nason2026}. In \Pythia, three threshold models that simulate the bound-state effects on top of the Born-level $t\bar{t}$ process are provided. These options also allow one to choose between an approach using Coulomb factors only and one based on the Green's function, and one can also set top-quark width and Green's-function width separately~\cite{Sjostrand2026}. In \Powheg, the models are implemented at next-to-leading-order (NLO) accuracy within the \hvq~\cite{Frixione2007_2} and \bbfourl~\cite{Jezo2016} generators, following the approach of Refs.~\cite{Nason2025,Nason2026}. In the new prescriptions of the \Powheg's \hvq and \bbfourl method, the NLO predictions that include the threshold effects avoid the potential double counting problem which affects the Green's function approach at higher orders. In their respective papers, predictions of  \Pythia~\cite{Sjostrand2026} models and \hvq~\cite{Nason2026} models were also compared with those of Ref.~\cite{Garzelli2026} and found to be compatible with theoretical predictions obtained for the top quark pair threshold. 

In this work, the new generators (or prescriptions) introduced in Refs.~\cite{Fuks2025,Sjostrand2026,Nason2026} are compared with published measurements of the top-quark pair production in the dilepton decay channel~\cite{CMS2019,ATLAS2025}, using particle-level observables. The analyses used here are publicly available as \Rivet routines~\cite{Rivet4} in \Rivet-4.1.3. Each model is passed through the \Rivet routines and a $\chi^2$ value is evaluated to quantify the relative improvement offered by each model compared with the baseline model for a set of observables sensitive to the excess. For the \Powheg models, which are available at NLO accuracy in QCD, a comparison with the same models implemented at LO accuracy is also included. In addition, a \MiNNLO sample~\cite{Monni2020,Mazzitelli2021,Mazzitelli2022} is generated, in which the next-to-next-to-leading-order (NNLO) QCD calculation of top-quark pair production is matched to the parton shower. This sample provides the state-of-the-art QCD predictions without any ``toponium'' contribution and therefore sets the baseline against which the threshold excess in the data is assessed.

This paper is organised as follows. \autoref{sec:setup} describes the Monte Carlo samples and the software setup used to produce them. \autoref{sec:analysis} presents the particle-level observables and the $\chi^2$ procedure used to quantify the agreement between each generator and the data, and discusses the results of the comparison. \autoref{sec:conclusions} presents the conclusions.

\section{Monte Carlo event generator setup}\label{sec:setup}

As noted above, in this study multiple Monte Carlo (MC) generator prescriptions are used to compare the MC predictions for the top-quark pair threshold with the corresponding ATLAS and CMS measurements which are implemented as \Rivet routines~\cite{Rivet4}. For all studies, the dilepton decay channel is considered, $t\bar{t}\to W^{+}b\,W^{-}\bar{b}\to \ell^{+}\nu_{\ell}\,b\,\ell^{-}\bar{\nu}_{\ell}\,\bar{b}$, with $\ell\in\{e,\mu\}$, at the LHC Run~2 energy $\sqrt{s}=13\,\TeV$.

The simulation of toponium formation in the \MG~\cite{MG5AMC}  event generator follows the prescription set out in Ref.~\cite{Fuks2025}. This method simulates the gluon-gluon fusion process and models the formation of toponium in the colour-singlet channel at LO accuracy in QCD. The matrix elements of $gg\to t\bar{t}$ are first generated using the standard \MG procedure and then reweighted using an NRQCD-inspired Green's function. This procedure simulates only an isolated toponium signal. To combine it with the standard $t\bar{t}$ process, the latter is additionally simulated at LO accuracy without toponium formation, and the decays of the top quarks are generated by \MS~\cite{Artoisenet2013} thereby retaining spin correlations. As shown in Eq.~(16) of Ref.~\cite{Fuks2025}, the double counting is cancelled during the toponium reweighting at LO. The simulations use the CT18NLO parton distribution function (PDF) set~\cite{CT18}, accessed through the \LHAPDF software~\cite{Buckley2015} . The top-quark mass ($m_t$) and decay width ($\Gamma_t$) used in this modelling were set to $m_t=173\,\gev$ and $\Gamma_t=1.49\,\gev$, respectively. Finally, the LO simulations are matched to a parton shower generated with \Pythia~\cite{PYTHIA83} (version 8.318) using the ATLAS A14 tune~\cite{ATLASA14}.\footnote{For all studies in this paper, Multi-parton Interactions were disabled.}

Ref.~\cite{Sjostrand2026} introduces new models implemented within the \Pythia framework , which provides the following models accessed via the \texttt{TopThreshold} option: \texttt{model0}, the Born-level $t\bar{t}$ process, \texttt{model1}, an alternative that simulates effects at the top-quark pair threshold, such as Coulomb factors, and in which double counting is accounted, and finally, \texttt{model2} and \texttt{model3}, which use the Green's function formalism. In \texttt{model2}, it is possible to set the top-quark width and the Green's-function width, both of which are set to $1.34\,\gev$ here. In addition to the widths, \texttt{model3} subtracts the Green's-function width from the top-quark decay width. These models also allow one to set the spin correlations according to the pseudoscalar nature of $t\bar{t}$.\footnote{The \Pythia implementation and its threshold options are documented in the online manual, \url{https://pythia8.web.cern.ch/manuals/pythia8318/}.} This option is included in the generation for all four models, including \texttt{model0}, so that the additional effect of the spin-correlation component is retained. This effect is assumed to hold up to $10\,\gev$ above the threshold and then scales down linearly to $20\,\gev$ above the threshold. However, the nominal $t\bar{t}$ decays are implemented  without angular correlations for these models. The models also consider the effects due to the colour singlet and octet cases for the $gg\to t\bar{t}$ and $q\bar{q} \to t\bar{t}$ processes. The colour-singlet fraction is set to $2/7$ for $gg\to t\bar{t}$ and to zero for $q\bar{q}\to t\bar{t}$. In these simulations the top-quark mass is set to $m_t=172.5\,\gev$ and the nominal top-quark width to $\Gamma_t=1.34\,\gev$ for all models; in \texttt{model3} the top-quark width entering the decay is reduced to $\Gamma_t=1.24\,\gev$, with a Green's-function width of $0.10\,\gev$, so that the two sum to the nominal value. This follows the recommendation in Ref.~\cite{Sjostrand2026}, based on typical usage of the code. As before, the simulations use the ATLAS A14 tune~\cite{ATLASA14}. The default PDF set in \Pythia, \texttt{NNPDF23\_lo\_as\_0130\_qed}~\cite{Ball2013}, is used.

In the \Powheg procedure, the newly introduced models capture the threshold effects at higher orders, in particular at NLO in QCD. The models in Ref.~\cite{Nason2026} are based on the \hvq~\cite{Frixione2007_2} and \bbfourl~\cite{Jezo2016} generators. The nominal \hvq generator describes heavy-flavour pair production with NLO matrix elements matched to the parton shower (PS). The nominal \bbfourl generator instead describes the full $b\bar{b}\,4\ell$ final state at NLO+PS, including the $Wt$ single-top process and its interference with $t\bar{t}$ production. The new models introduced in Ref.~\cite{Nason2026} are dubbed \texttt{thr1} and \texttt{thr2} within the \hvq framework, while the \bbfourl model exists as a standalone modified generator implementing threshold corrections. The \texttt{thr1} model implements threshold corrections in the limit $\Gamma_t \to 0$, whereas \texttt{thr2} includes finite width as well as running-width effects as per the prescription~\cite{Nason2026}. The top-quark mass was set to $m_t=172.5\,\gev$ for both the \hvq and \bbfourl generators, and the top-quark decay width in \hvq to $\Gamma_t=1.31\,\gev$. The \bbfourl generator treats the $t\bar{t}$ system fully off-shell. All \Powheg samples use the NNPDF3.0 NLO PDF set~\cite{NNPDF30} via \LHAPDF~\cite{Buckley2015}. Both colour-singlet and colour-octet contributions were retained.  For \texttt{thr2} the \texttt{widths\_corr} option was set to 1. In these generators, the angular correlations between the top quarks are taken into account. For \hvq or \bbfourl generator, the events simulating the baseline $t\bar{t}$ process without threshold corrections were also generated. The \texttt{thr1} and \texttt{thr2} samples were also generated for the \hvq model, as was the \bbfourl-thr generator, referred to here as the \bbfourl generator with threshold corrections. Additional samples were generated in which only the Born term and the bound-state ($\delta$-function) contribution are kept. The simulations were carried out at both LO and NLO accuracy in QCD and matched to a parton shower generated with \Pythia~\cite{PYTHIA83}  using the ATLAS A14 tune~\cite{ATLASA14}.

A \MiNNLO sample was also generated using the \texttt{ttJ\_MiNNLO} process of \Powheg, which provides NNLO+PS QCD predictions for top-quark pair production~\cite{Monni2020,Mazzitelli2021,Mazzitelli2022}. This sample is included as a reference, in particular to assess how well the current state-of-the-art NNLO+PS description reproduces the observables sensitive to the threshold in the absence of any toponium modelling. Because it contains no bound-state contribution, this sample tests to what extent the particle-level observables considered here can be described by higher-order QCD corrections alone, without an additional threshold contribution. The dilepton decay is generated by \Powheg at the matrix-element level, including spin correlations. The events are showered with \Pythia  using the ATLAS A14 tune~\cite{ATLASA14}. In this sample, the top-quark mass is set to $m_t=173.3\,\gev$ and the top-quark width entering the generation is set to $\Gamma_t=1.35\,\gev$, and the NNPDF3.1 NNLO PDF set~\cite{NNPDF31} is used.

The generator setups are summarised in \autoref{tab:generator_setup}. This paper investigates whether the newer generators introduced in Refs.~\cite{Sjostrand2026,Nason2026} are better than the conventional $t\bar{t}$ generators not implementing bound-state effects at explaining the excess of events. As noted from the simulation setup discussed here, the fundamental parameters such as the top-quark mass and top-quark width, which play an important role at the top-quark pair threshold, are chosen differently in the models. This parameter set is taken from the papers introducing the generators or the suggested run card(s) included in the software repository. The QCD accuracy differs between models. Moreover, it is known that the \Pythia generator does not incorporate spin correlations for conventional $t\bar{t}$ samples, which are important for the $t\bar{t}$ angular distributions. A comparison of which generator is better is thus postponed to future work, when similar parameters are chosen.

\begin{table}[htbp]
\centering\small
\caption{Monte Carlo generators and physics inputs for the dilepton samples
of \autoref{sec:setup}. For each setup, the table lists the generator, the process type, the QCD order
at which the sample is generated and matched to the parton shower (PS), the
top-quark mass ($m_t$) and decay width ($\Gamma_t$) used in the simulation,
and the PDF set employed. \texttt{Born+$\delta$}/``full''  retain only the Born
term and the bound-state ($\delta$-function) contribution, or the complete threshold
correction, respectively.}
\label{tab:generator_setup}
\fittable{%
\begin{tabular}{llllll}
\toprule
Generator & Process Type & QCD Order & $m_t$ [GeV] & $\Gamma_t$ [GeV] & PDF \\
\midrule
\MG & continuum baseline & LO+PS & 173.0 & 1.49 & CT18NLO~\cite{CT18} \\
\MG & colour-singlet component & LO+PS & 173.0 & 1.49 & CT18NLO \\
\MG & continuum+colour-singlet & LO+PS & 173.0 & 1.49 & CT18NLO \\
\Powheg\ \texttt{hvq} & baseline & LO+PS & 172.5 & 1.31 & NNPDF3.0 NLO~\cite{NNPDF30} \\
\Powheg\ \texttt{hvq} & baseline & NLO+PS & 172.5 & 1.31 & NNPDF3.0 NLO \\
\Powheg\ \texttt{hvq} & thr1, full & LO+PS & 172.5 & 1.31 & NNPDF3.0 NLO \\
\Powheg\ \texttt{hvq} & thr1, Born+$\delta$ & NLO+PS & 172.5 & 1.31 & NNPDF3.0 NLO \\
\Powheg\ \texttt{hvq} & thr1, full & NLO+PS & 172.5 & 1.31 & NNPDF3.0 NLO \\
\Powheg\ \texttt{hvq} & thr2, full & LO+PS & 172.5 & 1.31 & NNPDF3.0 NLO \\
\Powheg\ \texttt{hvq} & thr2, Born+$\delta$ & NLO+PS & 172.5 & 1.31 & NNPDF3.0 NLO \\
\Powheg\ \texttt{hvq} & thr2, full & NLO+PS & 172.5 & 1.31 & NNPDF3.0 NLO \\
\Powheg\ \texttt{bb4l} & baseline & LO+PS & 172.5 & {--} & NNPDF3.0 NLO \\
\Powheg\ \texttt{bb4l} & baseline & NLO+PS & 172.5 & {--} & NNPDF3.0 NLO \\
\Powheg\ \texttt{bb4l} & Born+$\delta$ & LO+PS & 172.5 & {--} & NNPDF3.0 NLO \\
\Powheg\ \texttt{bb4l} & full & LO+PS & 172.5 & {--} & NNPDF3.0 NLO \\
\Powheg\ \texttt{bb4l} & Born+$\delta$ & NLO+PS & 172.5 & {--} & NNPDF3.0 NLO \\
\Powheg\ \texttt{bb4l} & full & NLO+PS & 172.5 & {--} & NNPDF3.0 NLO \\
\Pythia & \texttt{model0}, baseline & LO+PS & 172.5 & 1.34 & NNPDF2.3 LO (QCD+QED)~\cite{Ball2013} \\
\Pythia & \texttt{model1} & LO+PS & 172.5 & 1.34 & NNPDF2.3 LO (QCD+QED) \\
\Pythia & \texttt{model2} & LO+PS & 172.5 & 1.34 & NNPDF2.3 LO (QCD+QED) \\
\Pythia & \texttt{model3} & LO+PS & 172.5 & 1.24 ; $\Gamma_t^{\mathrm{Green}}=0.10$ & NNPDF2.3 LO (QCD+QED)\\
\Powheg\ \MiNNLO & baseline & NNLO+PS & 173.3 & 1.35 & NNPDF3.1 NNLO~\cite{NNPDF31} \\
\bottomrule
\end{tabular}}
\end{table}

\section{Analysis}\label{sec:analysis}

The simulated events are analysed by passing them through the \Rivet implementations of Refs.~\cite{ATLAS2025,CMS2019} corresponding to top-quark pair production in the dilepton decay channel from the ATLAS and CMS collaborations, respectively. The more recent ATLAS analysis considers a luminosity of $140\,\mathrm{fb}^{-1}$, whereas the CMS analysis corresponds to a luminosity of $35.9\,\mathrm{fb}^{-1}$. The various analyses are provided as public \Rivet routines in the \Rivet software version 4.1.3, which was used for this analysis, built with an interface to \textsc{HepMC3}~\cite{Buckley2021} for the event record, \FastJet~\cite{Cacciari2012} for jet clustering, and \textsc{YODA}~\cite{Buckley2023} for histogramming. This analysis considers normalised differential distributions of particle-level measurements. In practice, the \Rivet routines corresponding to the tags \texttt{ATLAS\_2025\_I2971071} (dubbed here as \ATLASTwoFive)~\cite{ATLAS2025} and \texttt{CMS\_2018\_I1703993} (dubbed here as \CMSOneEight)~\cite{CMS2019} were used. The \ATLASTwoFive analysis provides unfolded measurements for $t\bar{t}\to e\mu$ and $t\bar{t}\to e\mu b\bar{b}$, where the former corresponds to standalone $t\bar{t}$, whereas the latter also includes single-top production. Therefore, the $t\bar{t}\to e\mu b\bar{b}$ channel can be used for comparison with the \bbfourl sample. In principle, it is possible to consider the $t\bar{t}\to e\mu b\bar{b}$ for other \Powheg samples at NLO, however, one additionally needs to consider a diagram removal technique for the $Wt$ process, and this also leads to an additional systematic uncertainty arising from the various diagram removal or subtraction algorithms that exist in practice (see, for example, Ref.~\cite{Frixione2019}). Therefore, in the present analysis, the \bbfourl sample is compared only with the $e\mu b\bar{b}$ selection of \ATLASTwoFive.

The $\chi^2$ test statistic used for the analysis, following Ref.~\cite{Fuks2026}, is defined as:
\begin{equation}
  \chi^2 = \mathrm{d}X_{n-1}^{\mathrm T}\, C_{n-1}^{-1}\, \mathrm{d}X_{n-1},
  \label{eq:chisq}
\end{equation}
where $\mathrm{d}X$ denotes the vector of differences between the predicted and measured bin contents and
$C=C_{\mathrm{exp}}+C_{\mathrm{pred}}$ is their covariance matrix. It is the sum of the covariance matrix provided by the experimental analyses with that of the MC predictions. In the latter case, this is a diagonal matrix with statistical uncertainty owing to the finite size of the MC sample. When the experimental covariance matrix is absent, a diagonal matrix of uncertainties is assumed instead. For the current analysis, theoretical systematic uncertainties from the MC predictions are ignored. A recent study assessing the systematic uncertainties such as the PDF, scale, and top-quark mass uncertainties for the top-quark pair threshold can be found in Ref.~\cite{Garzelli2026}. Finally, $n$ in the subscript refers to the number of bins ($N_{\mathrm{bins}}$) of a distribution. As normalised differential measurements are considered here, the normalisation reduces the number of degrees of freedom (the integral under the curve is unity for the given distribution). This is ensured by dropping the last bin from the measurements and taking $N_{\mathrm{dof}} = N_{\mathrm{bins}} - 1$, where $N_{\mathrm{bins}}$ represents the original number of bins in the \Rivet analysis.

The analysis presented here is carried out for a subset of particle-level measurements of variables which are sensitive to toponium formation. These variables were also used in previous studies~\cite{Fuks2026, ATLAS2025, ATLASThreshold2026} implemented with the $\eta_t$ model (the pseudoscalar toponium state) of Ref.~\cite{Fuks2025}. In particular, the list of variables from the \Rivet implementation of Ref.~\cite{ATLAS2025} (\ATLASTwoFive) is as follows. The one-dimensional variables are the invariant mass of the lepton pair ($m_{e\mu}$), the azimuthal angular difference between the electron ($e$) and muon ($\mu$), $\Delta \phi_{e\mu}$, and the absolute rapidity of the $e\mu$ system ($|y_{e\mu}|$). Moreover, two-dimensional distributions of $\Delta \phi_{e\mu}$, $|y_{e\mu}|$, and the lepton pseudorapidity ($|\eta^{\ell}|$) in bins of $m_{e\mu}$ were also considered. The $m_{e\mu}$ bins used are $(0,80)$, $(80,120)$, $(120,200)$, and $(200,500)\,\gev$. As for the variables from the \Rivet implementation of Ref.~\cite{CMS2019} (\CMSOneEight), the variables used are the invariant mass of the leptons $m_{\ell\ell}$, the azimuthal angular difference of the leptons $\Delta \phi_{\ell\ell}$, and the pseudorapidity difference of the leptons ($\Delta\eta_{\ell\ell}$).

\begin{figure}[htbp]
\centering
\includegraphics[width=0.49\linewidth]{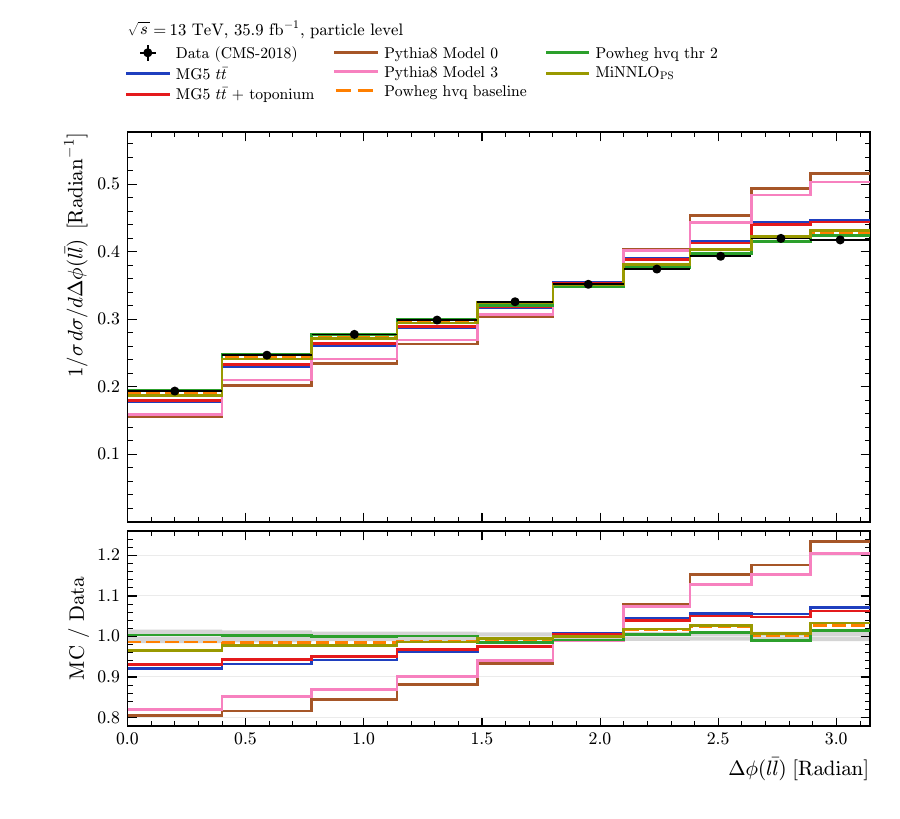}\hfill
\includegraphics[width=0.49\linewidth]{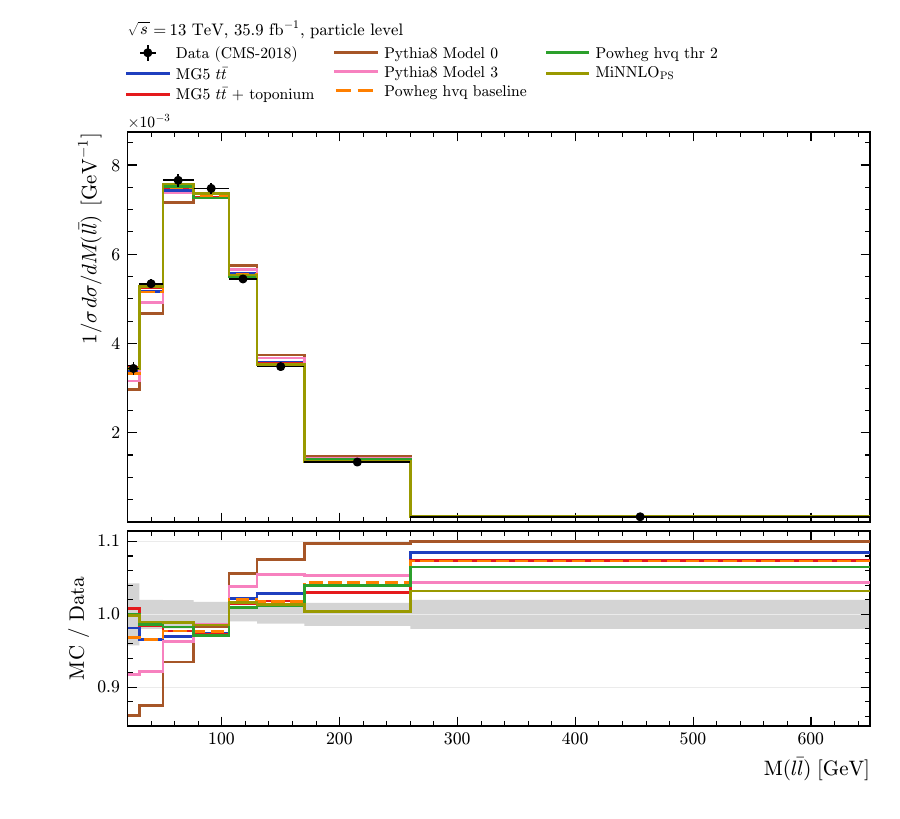}
\caption{Normalised particle-level distributions from the \CMSOneEight~\cite{CMS2019}
analysis ($35.9\,\mathrm{fb}^{-1}$) of the azimuthal angular difference between the
leptons, $\Delta\phi_{\ell\ell}$ (left), and the dilepton invariant mass,
$m_{\ell\ell}$ (right). For each generator (\MG, \Powheg\ \hvq, and
\Pythia), the baseline prediction without threshold effects is
compared with the corresponding threshold-improved setup of
\autoref{tab:generator_setup}, and the \MiNNLO prediction, which includes no
threshold contribution, is shown for reference; the $\chi^2/N_{\mathrm{dof}}$ values for these
observables are given in \autoref{tab:results-cms-1d}.}
\label{fig:curated-cms}
\end{figure}

The $\chi^2/N_{\mathrm{dof}}$ between the generator setups given in \autoref{tab:generator_setup} and the published particle-level normalised measurements are presented in \autoref{tab:results-cms-1d} and \autoref{tab:results-atlas25} for the variables that are sensitive to threshold effects. The corresponding distributions for different generator setups, comparing the baseline generator with the recommended option of the respective references are shown in \autoref{fig:curated-cms}.

\begin{table}[htbp]
\centering\scriptsize
\caption{Reduced $\chi^{2}$ ($\chi^{2}/N_{\mathrm{dof}}$) between each generator
setup of \autoref{tab:generator_setup} and the normalised particle-level
$t\bar{t}$ dilepton measurements of \CMSOneEight~\cite{CMS2019}
($35.9\,\mathrm{fb}^{-1}$), for the dilepton invariant mass ($m_{\ell\ell}$),
the azimuthal angular difference between the leptons ($\Delta\phi_{\ell\ell}$),
and the pseudorapidity difference between the leptons ($\Delta\eta_{\ell\ell}$).
The number of degrees of freedom
($N_{\mathrm{dof}}$) used in the $\chi^2$ evaluation are given under each
column.}
\label{tab:results-cms-1d}
\fittable{%
\begin{tabular}{lccc}
\toprule
Prediction & $m_{\ell\ell}$ & $\Delta\phi_{\ell\ell}$ & $\Delta\eta_{\ell\ell}$ \\
$N_{\mathrm{dof}}$ & 7 & 9 & 9 \\
\midrule
\MG\ continuum & 4.78 & 10.36 & 0.57 \\
\MG\ continuum$+$colour singlet & 3.15 & 7.85 & 0.73 \\
\midrule
\Powheg\ \texttt{hvq} baseline, LO+PS & 4.84 & 15.05 & 0.80 \\
\Powheg\ \texttt{hvq} baseline, NLO+PS & 4.13 & 1.33 & 0.28 \\
\Powheg\ \texttt{hvq} thr1, full, LO+PS & 1.37 & 6.22 & 1.94 \\
\Powheg\ \texttt{hvq} thr1, Born$+\delta$, NLO+PS & 3.07 & 1.14 & 0.26 \\
\Powheg\ \texttt{hvq} thr1, full, NLO+PS & 2.49 & 0.63 & 0.30 \\
\Powheg\ \texttt{hvq} thr2, full, LO+PS & 2.31 & 7.10 & 1.86 \\
\Powheg\ \texttt{hvq} thr2, Born$+\delta$, NLO+PS & 5.83 & 0.94 & 0.50 \\
\Powheg\ \texttt{hvq} thr2, full, NLO+PS & 3.03 & 0.55 & 0.36 \\
\midrule
\Pythia\ \texttt{model0} & 23.50 & 88.54 & 1.27 \\
\Pythia\ \texttt{model1} & 13.64 & 72.15 & 2.13 \\
\Pythia\ \texttt{model2} & 10.15 & 61.65 & 1.67 \\
\Pythia\ \texttt{model3} & 10.59 & 63.62 & 2.19 \\
\midrule
\Powheg\ \MiNNLO & 1.26 & 1.64 & 0.59 \\
\bottomrule
\end{tabular}}
\end{table}

\autoref{tab:results-cms-1d} gives the reduced $\chi^2$ ($\chi^2/N_{\mathrm{dof}}$) between the generators and the \CMSOneEight analysis, which uses $35.9\,\mathrm{fb}^{-1}$ of data. Comparing the baseline \MG generator against the \MG setup implemented in Ref.~\cite{Fuks2025}, one finds that the reduced $\chi^2$ improves for two of the three variables considered in this analysis, namely $m_{\ell\ell}$ and $\Delta \phi_{\ell\ell}$. The description of the $\Delta\eta_{\ell\ell}$ variable (see \autoref{fig:app-cms-d159}) offered by the generator setups at LO and NLO is not found to improve following the inclusion of threshold effects.

\begin{figure}[htbp]
\centering
\includegraphics[width=0.92\linewidth]{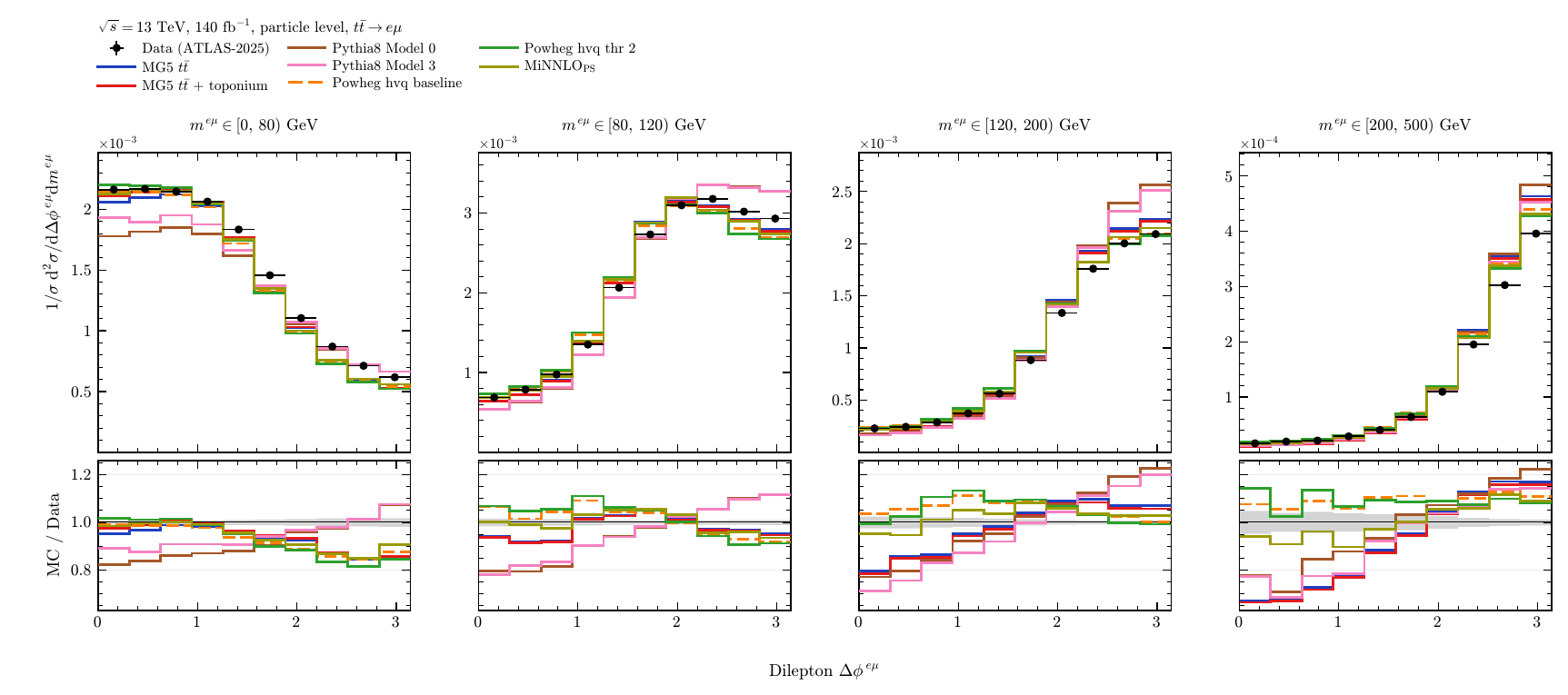}
\caption{Normalised two-dimensional \ATLASTwoFive~\cite{ATLAS2025}
($140\,\mathrm{fb}^{-1}$) distribution of the azimuthal angular difference
between the electron and muon, $\Delta\phi_{e\mu}$, unrolled in the four
$m_{e\mu}$ bins $(0,80)$, $(80,120)$, $(120,200)$, and $(200,500)\,\gev$,
for the $t\bar{t}\to e\mu$ selection. For each generator (\MG,
\Powheg\ \hvq, and \Pythia), the baseline prediction without
threshold effects is compared with the corresponding recommended
threshold-improved setup of \autoref{tab:generator_setup}, and the \MiNNLO
prediction, which includes no threshold contribution, is shown for reference; the
$\chi^2/N_{\mathrm{dof}}$ values for this observable are given in
\autoref{tab:results-atlas25}.}
\label{fig:curated-atlas25-2d}
\end{figure}

The corresponding distributions for different generator setups are shown in \autoref{fig:curated-atlas25-2d}.

\begin{table}[htbp]
\centering\scriptsize
\caption{Reduced $\chi^{2}$ ($\chi^{2}/N_{\mathrm{dof}}$) between each generator
setup of \autoref{tab:generator_setup} and the normalised particle-level
$t\bar{t}\to e\mu$ measurements of \ATLASTwoFive~\cite{ATLAS2025}
($140\,\mathrm{fb}^{-1}$), for the single-differential dilepton invariant mass
($m_{e\mu}$), azimuthal angular difference ($\Delta\phi_{e\mu}$), and absolute
rapidity ($|y_{e\mu}|$) of the $e\mu$ system, and for the corresponding
two-dimensional distributions of $\Delta\phi_{e\mu}$, $|y_{e\mu}|$, and the
lepton pseudorapidity $|\eta^{\ell}|$, each unrolled in the four $m_{e\mu}$
bins ($m_{e\mu}\times{}$ prefix; see \autoref{sec:analysis}). The \bbfourl
samples are compared with the $t\bar{t}\to e\mu b\bar{b}$ selection instead,
which additionally includes single-top production. The  number of degrees of freedom ($N_{\mathrm{dof}}$) used in the $\chi^2$
evaluation are given under each column. Horizontal rules separate the \MG,
\Powheg\ \texttt{hvq}, \Powheg\ \bbfourl, \Pythia, and \MiNNLO generator.}
\label{tab:results-atlas25}
\fittable{%
\begin{tabular}{lcccccc}
\toprule
& \multicolumn{3}{c}{Single-differential} & \multicolumn{3}{c}{Two-dimensional} \\
\cmidrule(lr){2-4}\cmidrule(lr){5-7}
Prediction & $m_{e\mu}$ & $\Delta\phi_{e\mu}$ & $|y_{e\mu}|$ & $m_{e\mu}\times\Delta\phi_{e\mu}$ & $m_{e\mu}\times|y_{e\mu}|$ & $m_{e\mu}\times|\eta^{\ell}|$ \\
$N_{\mathrm{dof}}$ & 14 & 19 & 8 & 39 & 19 & 35 \\
\midrule
\MG\ continuum & 7.08 & 9.48 & 3.94 & 21.28 & 10.86 & 7.21 \\
\MG\ continuum$+$colour singlet & 4.83 & 5.73 & 4.40 & 19.90 & 8.76 & 6.16 \\
\midrule
\Powheg\ \texttt{hvq} baseline, LO+PS & 8.63 & 17.28 & 1.79 & 33.15 & 10.78 & 10.73 \\
\Powheg\ \texttt{hvq} baseline, NLO+PS & 7.20 & 1.21 & 3.02 & 7.14 & 7.24 & 4.86 \\
\Powheg\ \texttt{hvq} thr1, full, LO+PS & 4.85 & 3.25 & 0.82 & 26.85 & 5.09 & 9.20 \\
\Powheg\ \texttt{hvq} thr1, Born$+\delta$, NLO+PS & 5.22 & 1.71 & 2.39 & 7.43 & 4.95 & 4.39 \\
\Powheg\ \texttt{hvq} thr1, full, NLO+PS & 4.05 & 3.13 & 2.15 & 8.19 & 3.37 & 3.85 \\
\Powheg\ \texttt{hvq} thr2, full, LO+PS & 4.81 & 4.89 & 1.47 & 28.94 & 6.23 & 10.18 \\
\Powheg\ \texttt{hvq} thr2, Born$+\delta$, NLO+PS & 8.51 & 2.33 & 3.12 & 9.66 & 6.82 & 5.82 \\
\Powheg\ \texttt{hvq} thr2, full, NLO+PS & 5.54 & 4.73 & 2.61 & 10.71 & 5.48 & 4.11 \\
\midrule
\Powheg\ \texttt{bb4l} baseline, LO+PS & 17.25 & 10.91 & 1.01 & 15.48 & 13.28 & 13.49 \\
\Powheg\ \texttt{bb4l} baseline, NLO+PS & 10.50 & 2.03 & 3.27 & 8.42 & 7.24 & 6.05 \\
\Powheg\ \texttt{bb4l} Born$+\delta$, LO+PS & 15.95 & 8.75 & 1.17 & 16.24 & 11.93 & 12.63 \\
\Powheg\ \texttt{bb4l} full, LO+PS & 6.38 & 1.47 & 1.43 & 13.87 & 3.54 & 7.43 \\
\Powheg\ \texttt{bb4l} Born$+\delta$, NLO+PS & 7.87 & 1.63 & 3.16 & 6.93 & 6.14 & 6.01 \\
\Powheg\ \texttt{bb4l} full, NLO+PS & 6.10 & 2.83 & 2.07 & 7.06 & 4.02 & 4.10 \\
\midrule
\Pythia\ \texttt{model0} & 45.51 & 134.56 & 55.29 & 77.90 & 89.64 & 33.12 \\
\Pythia\ \texttt{model1} & 29.27 & 102.41 & 56.06 & 62.39 & 75.48 & 25.70 \\
\Pythia\ \texttt{model2} & 20.63 & 87.53 & 60.80 & 53.67 & 66.73 & 22.59 \\
\Pythia\ \texttt{model3} & 23.87 & 90.12 & 63.75 & 57.29 & 73.13 & 22.41 \\
\midrule
\Powheg\ \MiNNLO & 4.93 & 1.14 & 3.33 & 7.36 & 3.67 & 3.76 \\
\bottomrule
\end{tabular}}
\end{table}

\autoref{tab:results-atlas25} gives the reduced $\chi^2$ ($\chi^2/N_{\mathrm{dof}}$) between the generators and the \ATLASTwoFive analysis, which uses $140\,\mathrm{fb}^{-1}$ of data. As observed for the \CMSOneEight analysis, replacing the baseline \MG continuum prediction with the \MG colour-singlet setup of Ref.~\cite{Fuks2025} improves the description of the single-differential $m_{e\mu}$ (see \autoref{fig:app-atlas25-d14}) and $\Delta\phi_{e\mu}$ distributions, whereas the $|y_{e\mu}|$ spectrum is not improved. Both the \texttt{thr1} and \texttt{thr2} prescriptions improve on the baseline \hvq prediction for $m_{e\mu}$ and for most of the two-dimensional observables. The \bbfourl sample, compared here with the $t\bar{t} \to e\mu b\bar{b}$ selection, improves on its baseline NLO+PS prediction for $m_{e\mu}$, and describes $\Delta\phi_{e\mu}$ well, once NLO corrections are included. In the \Pythia models, \texttt{model2} and \texttt{model3} lower the $m_{e\mu}$ and two-dimensional $\chi^2/N_{\mathrm{dof}}$ relative to \texttt{model0} and \texttt{model1}. In addition to the one-dimensional observables, the two-dimensional distributions $\Delta\phi_{e\mu}$, $|y_{e\mu}|$ (see \autoref{fig:app-atlas25-d25}), and the lepton pseudorapidity $|\eta^{\ell}|$ (see \autoref{fig:app-atlas25-d24}), each unrolled in bins of $m_{e\mu}$, are also considered. Since the lowest bin, $m_{e\mu} < 80\,\gev$, selects the region closest to the $t\bar{t}$ production threshold, these unrolled distributions isolate the phase space where toponium effects are largest and thereby add discriminating power between the generator setups.

Some of the per-generator distributions for the observables in \autoref{tab:results-cms-1d} and \autoref{tab:results-atlas25} that are not shown above are collected in Appendix~\ref{app:types}.

\section{Conclusions}\label{sec:conclusions}

In this paper, $t\bar{t}$ production near threshold was studied using some of the recent Monte Carlo event generators, and their predictions were compared with the ATLAS and CMS analyses implemented within \Rivet. The normalised differential distributions implemented in the \Rivet analyses labelled as \CMSOneEight~\cite{CMS2019} and \ATLASTwoFive~\cite{ATLAS2025} were compared with each generator setup. Broadly, this work finds that the inclusion of threshold effects introduced in the newer models of Refs.~\cite{Fuks2025, Sjostrand2026, Nason2026} for the \MG, \Pythia and \Powheg MC generators, respectively, tends to improve the reduced $\chi^2$ value, $\chi^2/N_{\mathrm{dof}}$, compared with the baseline models of the same generators available before these recent publications, although not for every observable. In particular, the description offered by the \MG model with toponium included is found to be better compared with the baseline \MG for most observables, the exceptions being $\Delta\eta_{\ell\ell}$ and $|y_{e\mu}|$. For \Powheg, the \hvq-\texttt{thr2} at NLO is found to offer a better description when compared with the baseline for most observables. \Powheg-\bbfourl-\texttt{thr} also shows improvements compared with baseline \bbfourl, except for $\Delta\phi_{e\mu}$. In the \Pythia models, \texttt{model3} tends to perform better compared with the baseline \texttt{model0}. The description offered by \MiNNLO is also found to have the lowest value for the reduced $\chi^2$ for some of the observables, although this is without the inclusion of the ``toponium effects''.

\appendix

\section{Per-generator distributions for the remaining observables}\label{app:types}

The figures below complement the curated comparisons of \autoref{sec:analysis} by
showing, one panel per generator (\MG, \Powheg\ \hvq, and \Pythia), some of the
observables that enter the $\chi^2/N_{\mathrm{dof}}$ tables of that section but are
not given in the main text. 

\begin{figure}[htbp]
\centering
\includegraphics[width=0.49\linewidth]{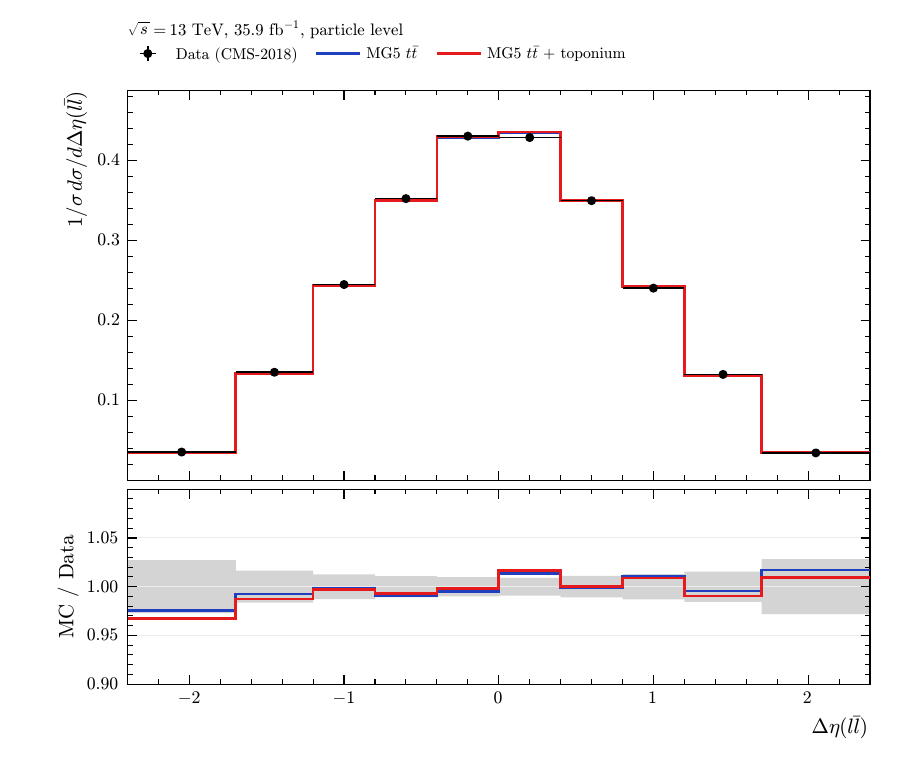}
\includegraphics[width=0.49\linewidth]{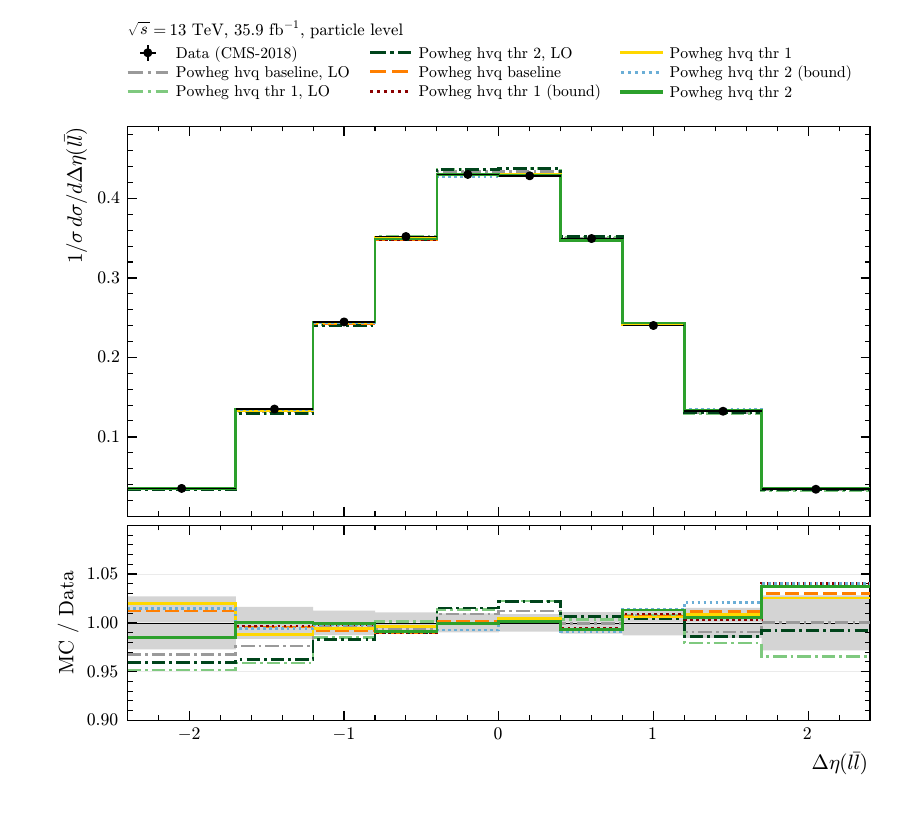}\\
\includegraphics[width=0.49\linewidth]{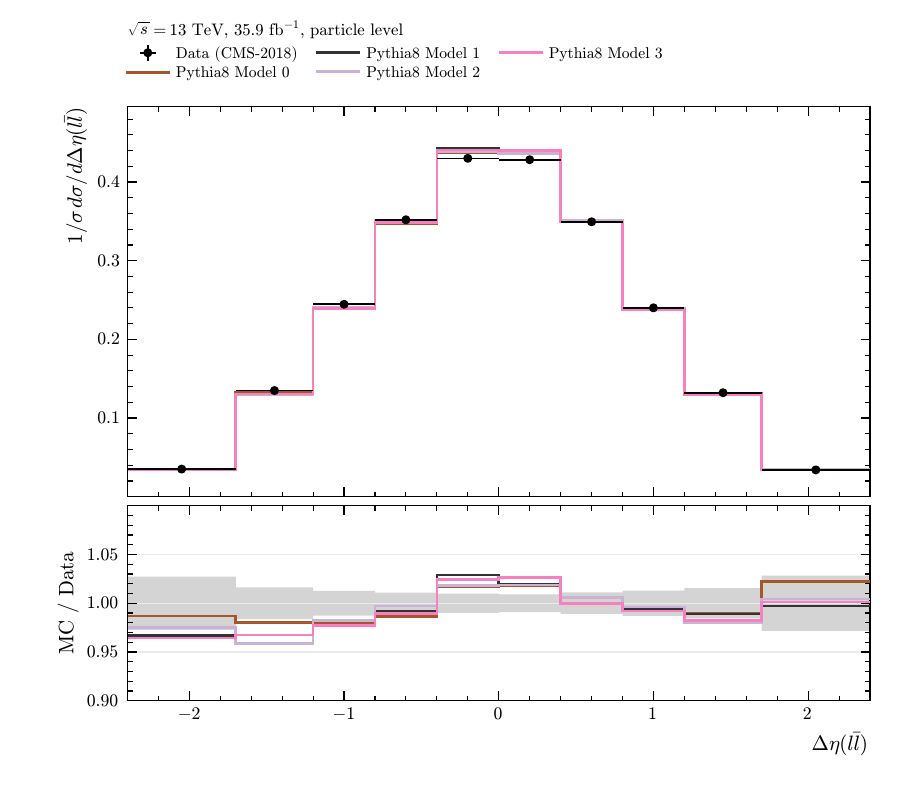}
\caption{Normalised \CMSOneEight~\cite{CMS2019} ($35.9\,\mathrm{fb}^{-1}$)
distribution of the pseudorapidity difference between the leptons,
$\Delta\eta_{\ell\ell}$, shown separately for the \MG (top left),
\Powheg\ \hvq (top right), and \Pythia (bottom). The corresponding
$\chi^2/N_{\mathrm{dof}}$ values are given in the $\Delta\eta_{\ell\ell}$
column of \autoref{tab:results-cms-1d}.}
\label{fig:app-cms-d159}
\end{figure}

\begin{figure}[htbp]
\centering
\includegraphics[width=0.49\linewidth]{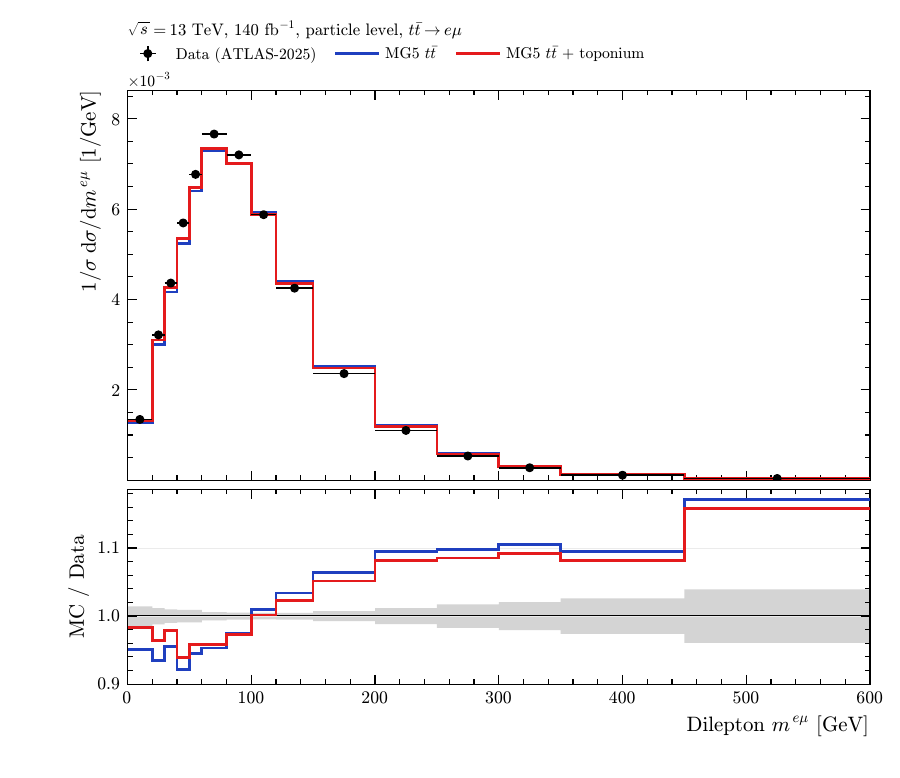}
\includegraphics[width=0.49\linewidth]{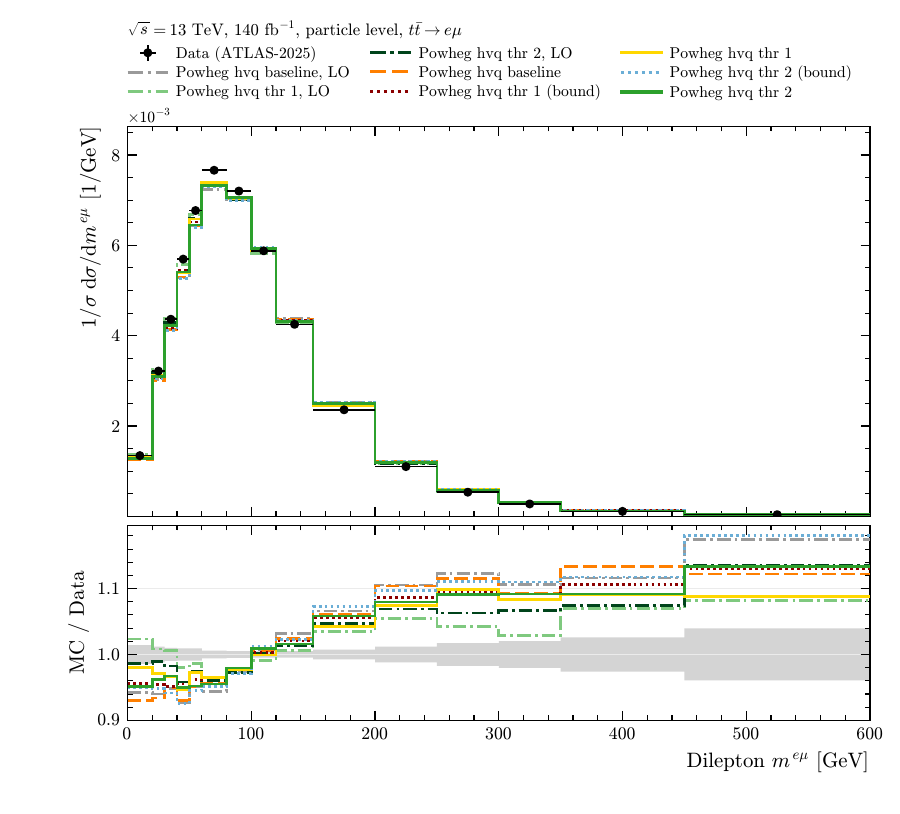}\\
\includegraphics[width=0.49\linewidth]{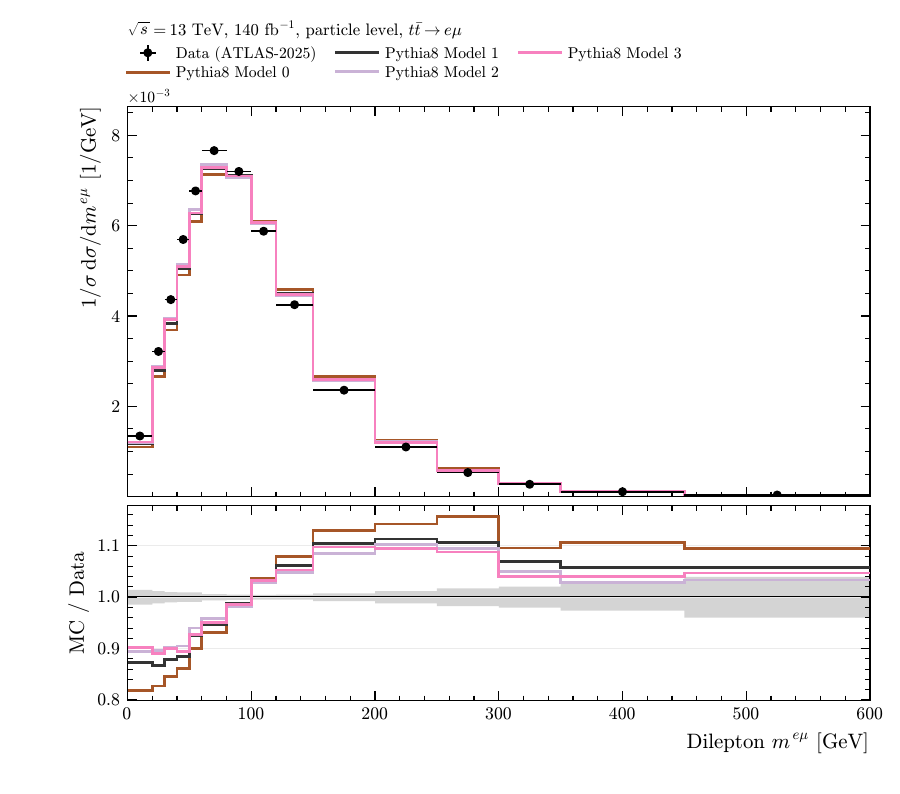}
\caption{Normalised single-differential \ATLASTwoFive~\cite{ATLAS2025}
($140\,\mathrm{fb}^{-1}$) dilepton invariant-mass distribution, $m_{e\mu}$,
for the $t\bar{t}\to e\mu$ selection, shown separately for the \MG (top
left), \Powheg\ \hvq (top right), and \Pythia (bottom). The
corresponding $\chi^2/N_{\mathrm{dof}}$ values are given in the $m_{e\mu}$
column of \autoref{tab:results-atlas25}.}
\label{fig:app-atlas25-d14}
\end{figure}

\begin{figure}[htbp]
\centering
\includegraphics[width=0.88\linewidth]{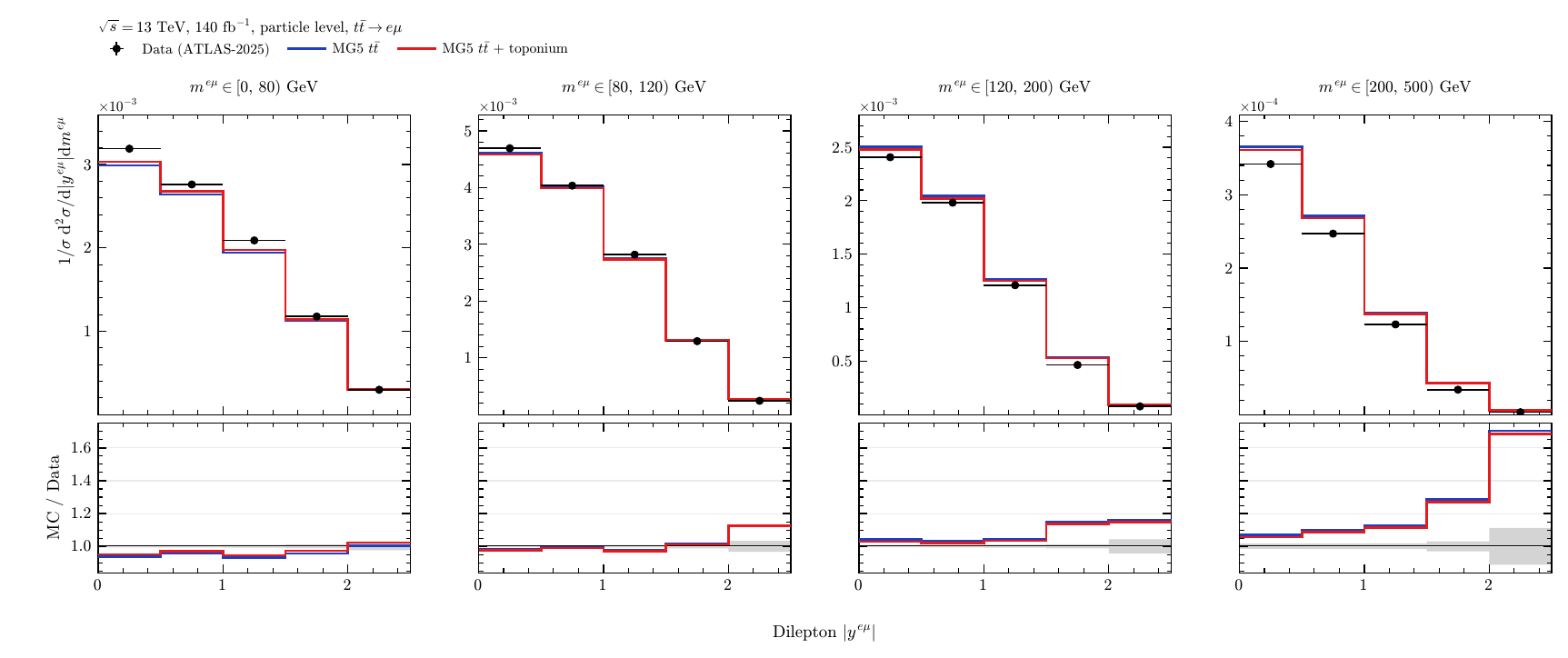}\\
\includegraphics[width=0.88\linewidth]{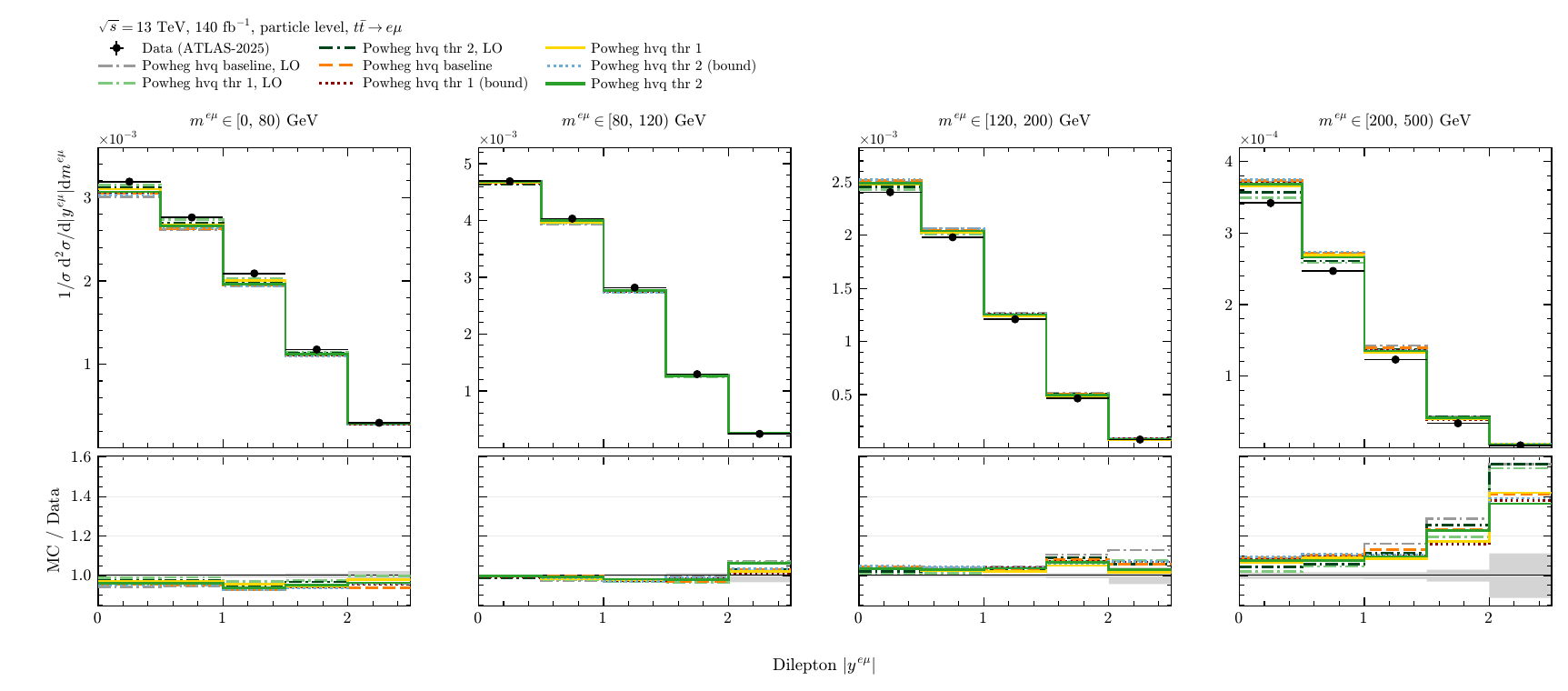}\\
\includegraphics[width=0.88\linewidth]{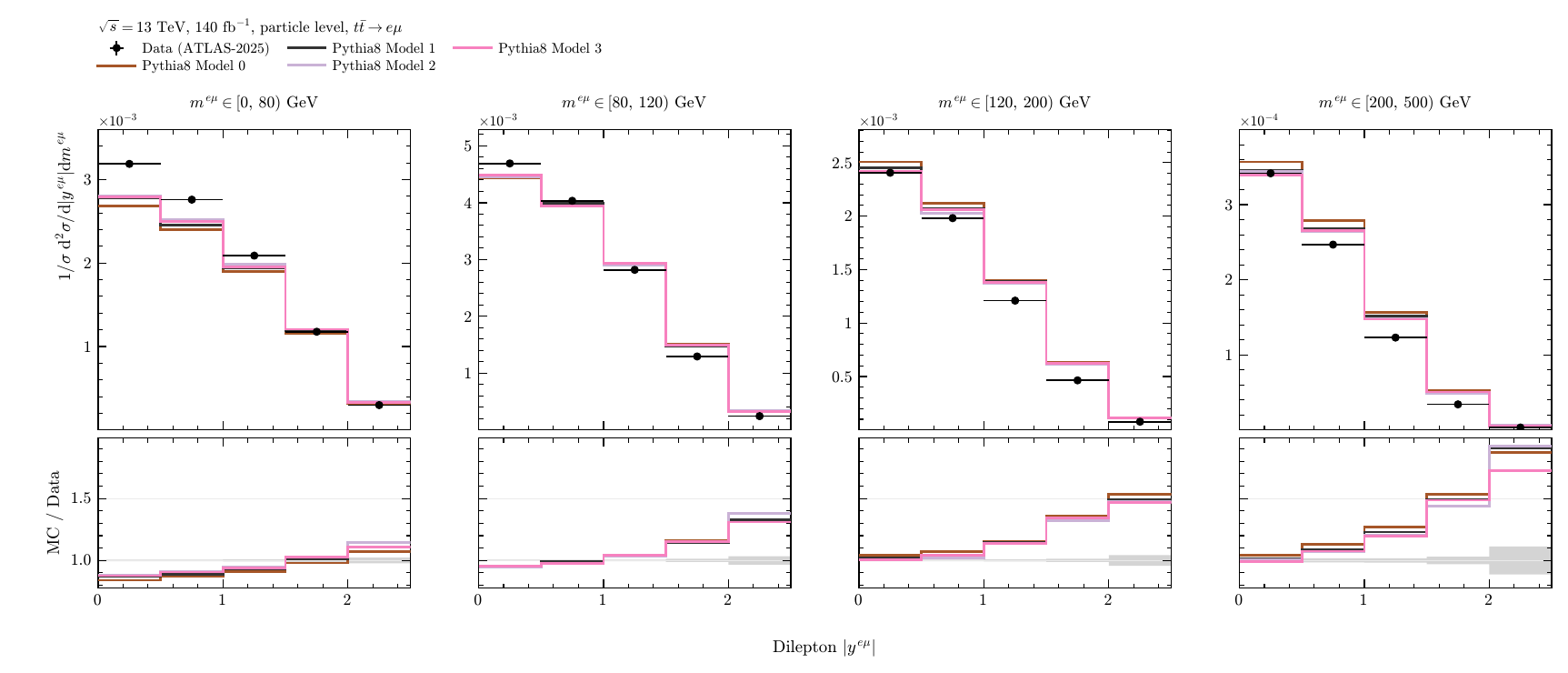}
\caption{Normalised two-dimensional \ATLASTwoFive~\cite{ATLAS2025}
($140\,\mathrm{fb}^{-1}$) distribution of the absolute rapidity of the $e\mu$
system, $|y_{e\mu}|$, unrolled in the four $m_{e\mu}$ bins $(0,80)$,
$(80,120)$, $(120,200)$, and $(200,500)\,\gev$, for the $t\bar{t}\to e\mu$
selection, shown separately for the \MG (top), \Powheg\ \hvq (middle), and
\Pythia (bottom). The corresponding $\chi^2/N_{\mathrm{dof}}$ values
are given in the $m_{e\mu}\times|y_{e\mu}|$ column of
\autoref{tab:results-atlas25}.}
\label{fig:app-atlas25-d25}
\end{figure}

\begin{figure}[htbp]
\centering
\includegraphics[width=0.88\linewidth]{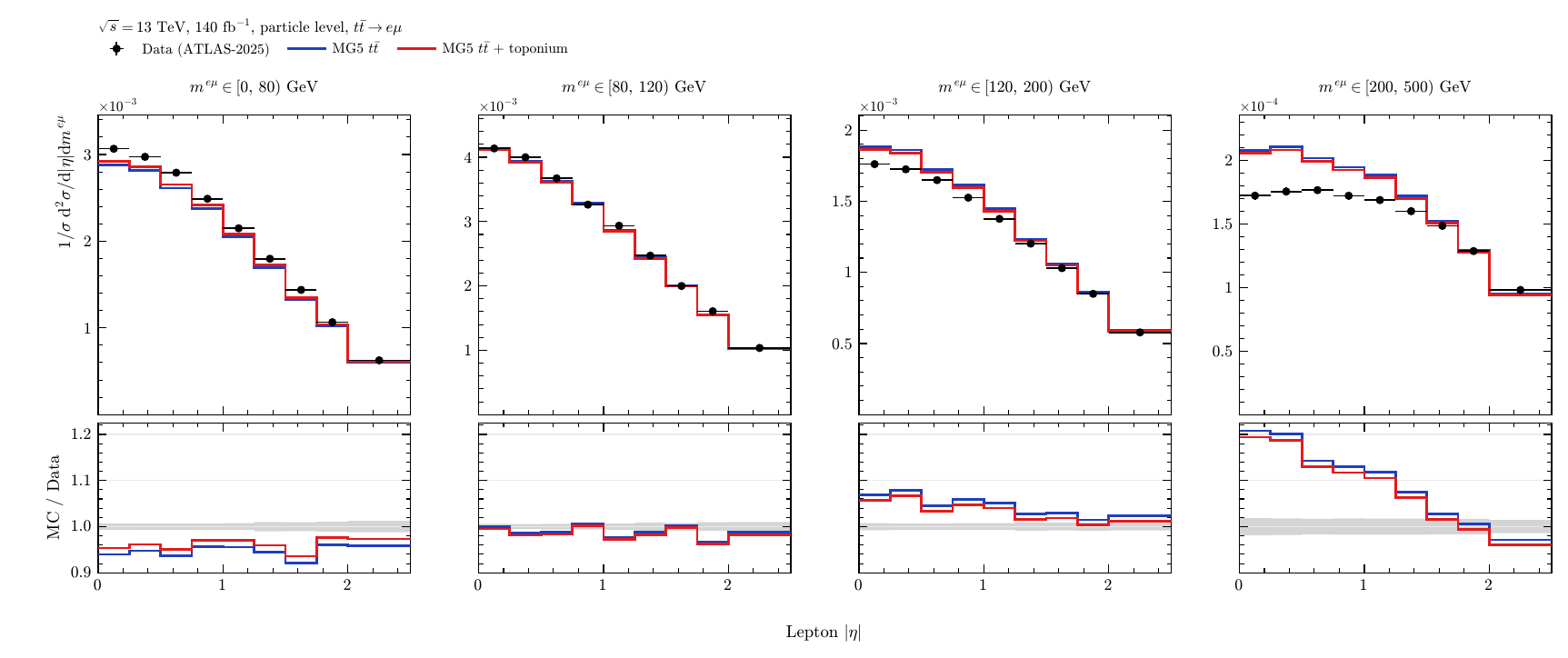}\\
\includegraphics[width=0.88\linewidth]{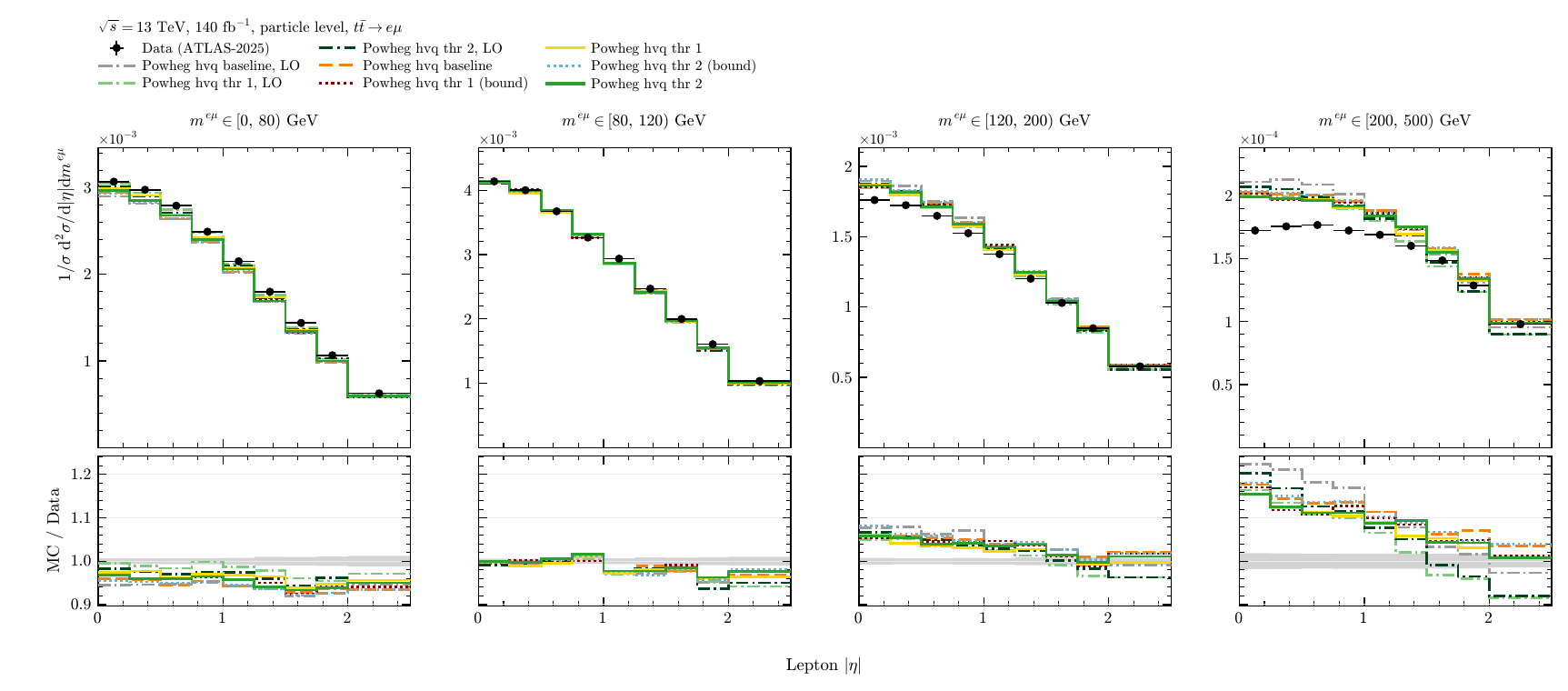}\\
\includegraphics[width=0.88\linewidth]{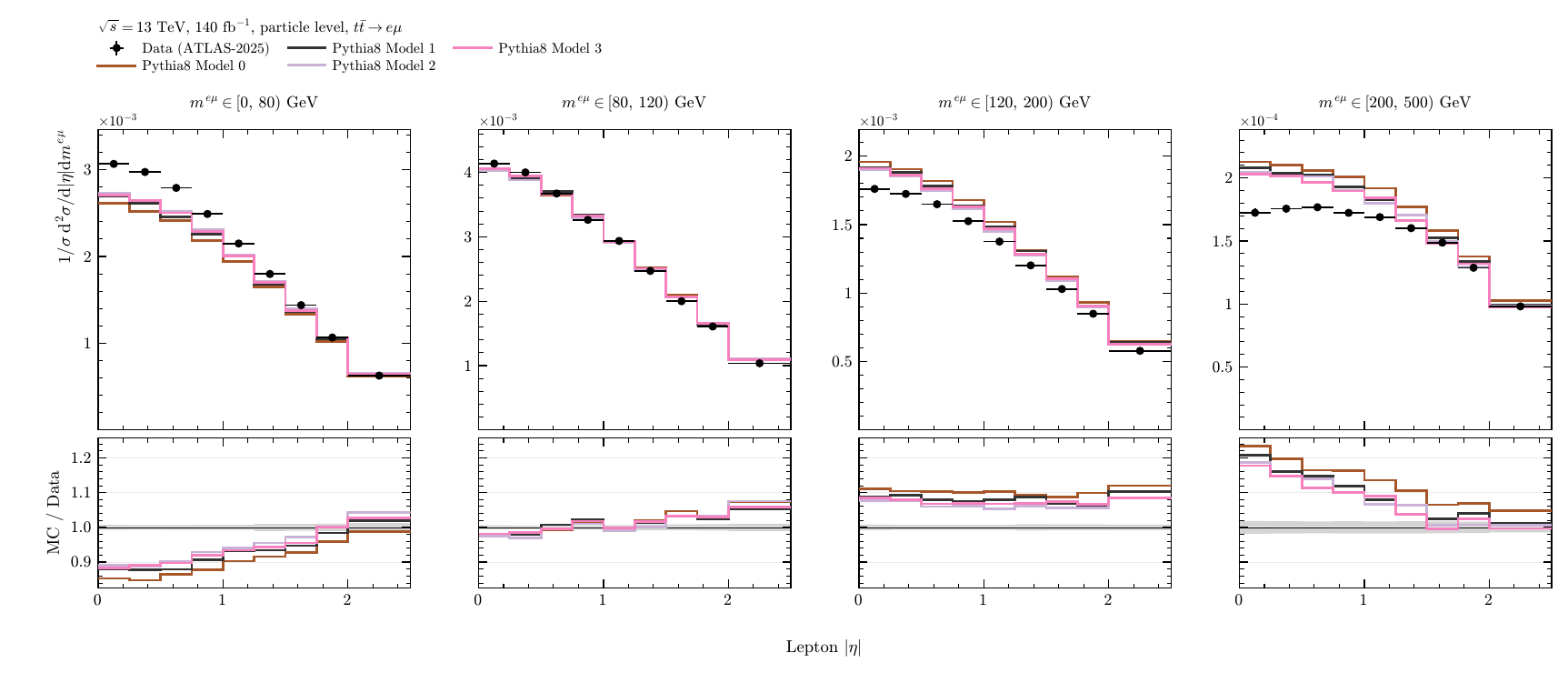}
\caption{Normalised two-dimensional \ATLASTwoFive~\cite{ATLAS2025}
($140\,\mathrm{fb}^{-1}$) distribution of the lepton pseudorapidity,
$|\eta^{\ell}|$, unrolled in the four $m_{e\mu}$ bins $(0,80)$, $(80,120)$,
$(120,200)$, and $(200,500)\,\gev$, for the $t\bar{t}\to e\mu$ selection,
shown separately for the \MG (top), \Powheg\ \hvq (middle), and \Pythia
(bottom). The corresponding $\chi^2/N_{\mathrm{dof}}$ values are
given in the $m_{e\mu}\times|\eta^{\ell}|$ column of
\autoref{tab:results-atlas25}.}
\label{fig:app-atlas25-d24}
\end{figure}

\clearpage \newpage


\bibliographystyle{JHEP}
\bibliography{biblio}

\end{document}